# Interband Tunneling in 2D Crystal Semiconductors

Nan Ma and Debdeep Jena[a]

*Department of Electrical Engineering, University of Notre Dame, Notre Dame, IN 46556*

Interband quantum tunneling of electrons in semiconductors is of intense recent interest as the underlying transport mechanism in tunneling field-effect transistors. Such transistors can potentially perform electronic switching with lower energy than their conventional counterparts. The recent emergence of 2-dimensional semiconducting crystals provides a new material platform for realizing such devices. In this work, we derive an analytical expression for understanding tunneling current flow in 2D crystal semiconductors. We apply the results to a range of 2D crystal semiconductors, and compare it with tunneling currents in 3D semiconductors. We also discuss the implications for tunneling devices.

a) Electronic mail: djena@nd.edu

2-dimensional (2D) crystals of graphene were first isolated in 2004.[1] The novel electronic and optical properties of graphene have been extensively studied since then.[2] Electronic transistors have been proposed with graphene for ultra-low power switching.[3-5] These proposed devices either exploit the symmetric zero-bandgap Dirac-cone bandstructure of graphene, or require the opening of energy bandgaps by quantum confinement. Soon after the isolation of graphene, *semiconducting* 2D crystals were reported.[6-8] 2D crystal semiconductors have been found in the transition metal dichalcogenide (TMD) material family, and the list is expected to expand in the future.[9] Taking advantage of an energy bandgap, ultrathin channels, and absence of broken bonds, conventional field-effect transistors (FETs) using 2D crystal semiconductors have shown high promise in initial investigations.[7,10]

A number of electronic switching devices have been proposed recently to address the power-dissipation problems of FETs.[11] Among the proposed devices, the tunneling FET (or TFET) has emerged as an attractive candidate. These devices take advantage of the potential of interband Zener-tunneling of electrons to beat the Boltzmann thermal limit of switching of 60 mV/decade. Initial experimental demonstrations show much promise.[12-14] The availability of 2D semiconducting crystals with bandgaps begs the question whether TFETs with attractive properties can be realized with them. Such 2D crystal TFETs, if realized, can take advantage of the ultrathin nature of the layers, and the absence of broken bonds to enable scaling of such devices to much smaller dimensions than 3D crystal semiconductors. A critical metric for TFETs is the on-state current, which is limited by interband tunneling of electrons. To date, interband tunneling in purely 2D semiconducting crystal junctions has not received sufficient attention, certainly not to the extent it has for 3D semiconductor p-n junctions since Zener's[15] and Esaki's works.[16] In this work, we take this first step. We derive an analytical expression for the tunneling current in 2D crystal semiconductors. The expression highlights the dependence of the tunneling current on the material parameters of the 2D crystal semiconductor such as its bandgap and effective masses. We apply the results to a range of 2D crystals, and discuss the implications for device applications.

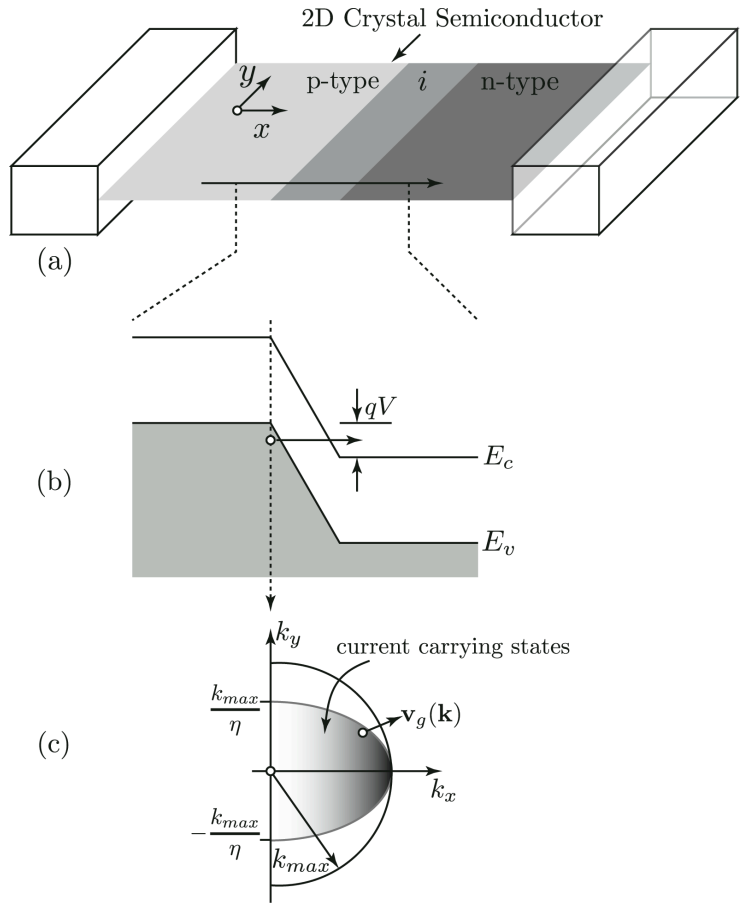

**Fig 1:** Schematic depiction of a 2D crystal *p-i-n* junction (a), the energy band-diagram (b), and the **k**-space distribution of current densities. The **k**-states contributing to interband tunneling current and the group velocity are indicated.

Consider the 2D crystal *p-i-n* junction shown schematically in Fig. 1 (a) with ohmic contacts to the *p*- and *n*-doped regions. The contacts would form the source and drain contacts of the corresponding TFET. We do not address the experimental challenges of doping and electrostatic gating in this work and focus exclusively on evaluating the 2-terminal tunneling current. Assume the doping in the *p*- and *n*-sides aligns the Fermi levels to the respective band-edges. Then, under no applied bias, $E_v^p = E_c^n$ and no net current flows across the junction. Here $E_v^p$ is the valence-band edge on the *p*-side, and $E_c^n$ is the conduction-band edge on the *n*-side. Under the application of a reverse bias voltage $V$, a finite energy window is created for electrons since $E_v^p - E_c^n = qV$. Within this energy window, electrons from the valence band can tunnel into the conduction band on the other side, as indicated in Fig 1 (b).

The electric current is obtained by summing the individual quantum-mechanical probability current contributions by each *k*-state electron, and multiplying it by $q$, the electron charge. The tunneling current is thus given by

$$I_T = q \frac{g_s g_v}{L_x} \sum_k v_g(k)(f_v - f_c) T_{wkb}, \qquad (1)$$

where $g_s = 2$ is the spin degeneracy, and $g_v$ is the valley degeneracy for 2D crystal single layers. Single-layer TMD 2D crystals have been found to have two valleys at the *K* and *K'* points of the Brillouin zone similar to 2D graphene[17] so we use $g_v = 2$. $L_x$ is the macroscopic length along the electric field (which will cancel out), $v_g(k) = \hbar^{-1} \nabla E(k)$ is the group velocity of carriers in the band $E(k)$, $f_v$, $f_c$ are the occupation functions of the valence and conduction bands respectively, and $T_{wkb}$ is the interband tunneling probability given by the Wentzel-Kramers-Brillouin (WKB) approximation. The sum is over all *k*-state electrons that are allowed to tunnel.

The tunneling probability $T_{wkb}$ is obtained by the WKB approximation in the following manner. For 2D crystals, electrons in the valence band of the *p*-side have a transverse kinetic energy $E_y = \hbar^2 k_y^2 / 2m_v^*$, where $\hbar k_y$ is the tranverse quasi-momentum, and $m_v^*$ is the valence band effective mass. The WKB tunneling probability is then given by[18]

$$T_{WKB} = \exp[-\frac{4\sqrt{2m_R^*}(E_g + E_y)^{3/2}}{3q\hbar F}] \approx T_0 \exp[-\frac{E_y}{\bar{E}}], \qquad (2)$$

where $T_0 = \exp[-4\sqrt{2m_R^*}E_g^{3/2}/3q\hbar F]$ is the tunneling probability of perpendicularly incident electrons, $\bar{E} = q\hbar F / 2\sqrt{2m_R^* E_g}$, $F$ is the (constant) electric field in the junction, and $m_R^*$ is the reduced effective mass given by $m_R^* = m_c^* m_v^* / (m_c^* + m_v^*)$. $m_v^*$, $m_c^*$ are the effective masses of electrons of the valence and conduction bands respectively. The above expression is found to be consistent with experimental results.[18] Note that the tunneling probability of electrons is lowered *exponentially* with their transverse kinetic energy as a consequence of lateral momentum conservation in the tunneling process. We neglect phonon emission or absorption processes here. To evaluate the tunneling current, we attach this tunneling probability to each electronic *k*-state, and sum it over all electrons incident on the tunneling barrier.

In Fig. 1, we concentrate on a particular 1D line as shown by the dashed line, at the *p-i* junction, which is the source side. Half of the electrons in the valence band in that line move to the right in the $+k_x$ direction, as indicated in the semi-circle in the *k*-space in Fig 1 (c). Since there are negligible electrons in the conduction band in that line, the current there must be carried by electrons in the valence band. Which of these right-going electrons are allowed to tunnel through the gap? In the absence of phonon scattering, tunneling is an

elastic process. This enforces the energy conservation requirement

$$E_v^p - \frac{\hbar^2}{2m_v^*}(k_{xp}^2 + k_{yp}^2) = E_c^n + \frac{\hbar^2}{2m_c^*}(k_{xn}^2 + k_{yn}^2), \qquad (3)$$

with the additional requirement that the lateral momentum be conserved, i.e., $k_{yp} = k_{yn} = k_y$. The energy and momentum conservation requirements thus lead to the relation

$$k_{xp}^2 + \frac{m_v^*}{m_R^*}k_y^2 = \frac{2m_v^*qV}{\hbar^2} - \frac{m_v^*}{m_c^*}k_{xn}^2. \qquad (4)$$

Let us define $k_{max}^2 = 2m_v^*qV/\hbar^2$ and $\eta^2 = m_v^*/m_R^*$. Note that $k_{max}$, the radius of the semi-circle in the $k$-space shown in Fig 1(c), is controlled by the applied voltage. Since there is an electric field in the $x$-direction, the momentum in that direction will not be conserved. For the electron to emerge on the right ($n$-) side, $k_{xn}$ must be non-zero, and thus $k_{xn}^2 \geq 0$, which implies

$$k_{xp}^2 + \eta^2 k_y^2 \leq k_{max}^2. \qquad (5)$$

The above condition defines a restricted elliptical area $A_T$ of the $k$-space semi-circle for electron states that are allowed to tunnel, as shown in Fig 1 (c). We can now evaluate the tunneling current for 2D semiconductor $p$-$i$-$n$ junctions. In the expression for the tunneling current (Eq. 1), the group velocity term is that of the valence band $k$-state $v_g(k) = \hbar k_x / m_v^*$. We skip the $p$-subscripts, since it is clear that the electrons tunnel from the valence band of the $p$-side. The expression for the tunneling current is then

$$I_T = q\frac{g_s g_v}{L_x} \sum_{(k_x,k_y)\in A_T} \frac{\hbar k_x}{m_v^*}(f_v - f_c)T_0 \exp[-\frac{\hbar^2 k_y^2}{2m_v^* \overline{E}}]. \qquad (6)$$

The sum over $k$-states is converted into an integral via the standard recipe $\sum_k (...) \to L_x L_y /(2\pi)^2 \times \int dk_x dk_y (...)$. Due to the 'filtering' brought about by the requirements of energy and momentum conservation (Eq. 5), the $k$-space integral is evaluated over the restricted area $A_T$. The tunneling current *per unit width*, or the *current density* is then given by

$$J_T^{2D} = \frac{I_T}{L_y} = \frac{q g_s g_v \hbar T_0}{(2\pi)^2 m_v^*} \int_{-k_{max}/\eta}^{+k_{max}/\eta} dk_y \exp[-\frac{\hbar^2 k_y^2}{2m_v^* \overline{E}}] \int_0^{\sqrt{k_{max}^2 - \eta^2 k_y^2}} dk_x k_x (f_v - f_c). \tag{7}$$

Electrons incident normal to the junction have no transverse momentum, and carry most of the tunneling current. The number of electron states allowed to tunnel reduces as their transverse directed momentum increases, as shown schematically in Fig.1 (c). The current carried by these states with transverse momentum is further damped by the $\exp[-E_y/\overline{E}]$ factor, leading to further filtering and momentum collimation. The maximum tunneling current is carried by states closest to $(k_x, k_y) = (k_{max}, 0)$ as indicated by the shading in the figure.

To evaluate the current, the integral in *k*-space should be evaluated. At $T \to 0$ K, $f_v - f_c \approx 1$ for the energy window of current-carrying electrons. This relation remains an excellent approximation at room temperature. The interband tunneling current per unit width (μA/μm) in a 2D crystal *p-i-n* junction then evaluates to

$$J_T^{2D} = \frac{q^2}{h} (\frac{g_s g_v T_0}{2\pi}) \sqrt{\frac{2m_v^* \overline{E}}{\hbar^2}} \times [\sqrt{\pi}(V - \frac{V_0}{2}) \text{Erf}(\sqrt{\frac{V}{V_0}}) + \sqrt{\frac{V}{V_0}} \exp(-\frac{V}{V_0})], \tag{8}$$

where Erf[...] stands for the error function, and we have defined $V_0 = \eta^2 \overline{E}/q$. Eq. 8 is the central result of this work. The expression shows the dependences on various bandstructure and junction parameters explicitly. For small reverse bias voltages $V << V_0$, the tunneling current varies as $J_T^{2D} \sim V^{3/2}$ to leading order. This is consistent with a recent report investigating dimensionality effects on tunneling.[19] For larger voltages when $V >> V_0$, Erf[...] $\to 1$, and we get a linear dependence of the tunneling current on the voltage

$$J_T^{2D} \approx \left[\frac{q^2}{h}(\frac{g_s g_v}{2\pi})\sqrt{\frac{2\pi m_v^*}{\hbar^2} \cdot \frac{q\hbar F}{\sqrt{8m_R^* E_g}}} \times T_0\right] V, \qquad (9)$$

where the Landauer quantum of conductance is split off. The entire square root term is an effective wavevector with units of inverse length, leading to units of current per unit width. The corresponding current density for tunneling current in 3-dimensional semiconductors is given by $J_T^{3D} \approx \left[\frac{q^2}{h}(\frac{g_s g_v}{2\pi})(\sqrt{\frac{2m_R^* E_g}{\hbar^2}}\frac{qF}{E_g}) \times T_0\right] V$ with units of current per unit area. The WKB term is similar for 2D and 3D crystals. The prefactor for 3D semiconductors goes as $\sim F\sqrt{m^*/E_g}$, whereas for 2D crystals it goes as its square root $\sim \sqrt{F\sqrt{m^*/E_g}}$. In quasi-2D systems multiple subbands may be involved in transport. Then we sum the current from each subband with the respective band parameters. When the temperature $T$ is high, the assumption $f_v - f_c \approx 1$ may no longer be suitable. In that case, the 2D tunneling current becomes

$$J_T^{2D} = \frac{qg_s g_v T_0 k_B T}{(2\pi)^2 \hbar} \int_{-k_{max}/\eta}^{+k_{max}/\eta} \ln\left\{\frac{(e^{\beta(qV-\eta^2 E_y)} + e^{-\beta E_y})(1+e^{\beta(qV-E_y)})}{(e^{\beta(qV-\eta^2 E_y)} + e^{\beta(qV-E_y)})(1+e^{-\beta E_y})}\right\} \exp[-\frac{E_y}{\bar{E}}] dk_y, \qquad (10)$$

where $\beta = 1/k_B T$ and $k_B$ is the Boltzmann constant. We have not simplified this expression analytically, but the numerical evaluation is discussed. The interband tunneling current densities of various 2D crystals at $T$=4 K and 300 K are plotted as solid and dashed lines in Fig. 2 (a) respectively. As is evident, the temperature dependence is rather weak. The material constants (bandgaps and effective masses) are obtained from Ref. 17, 20.

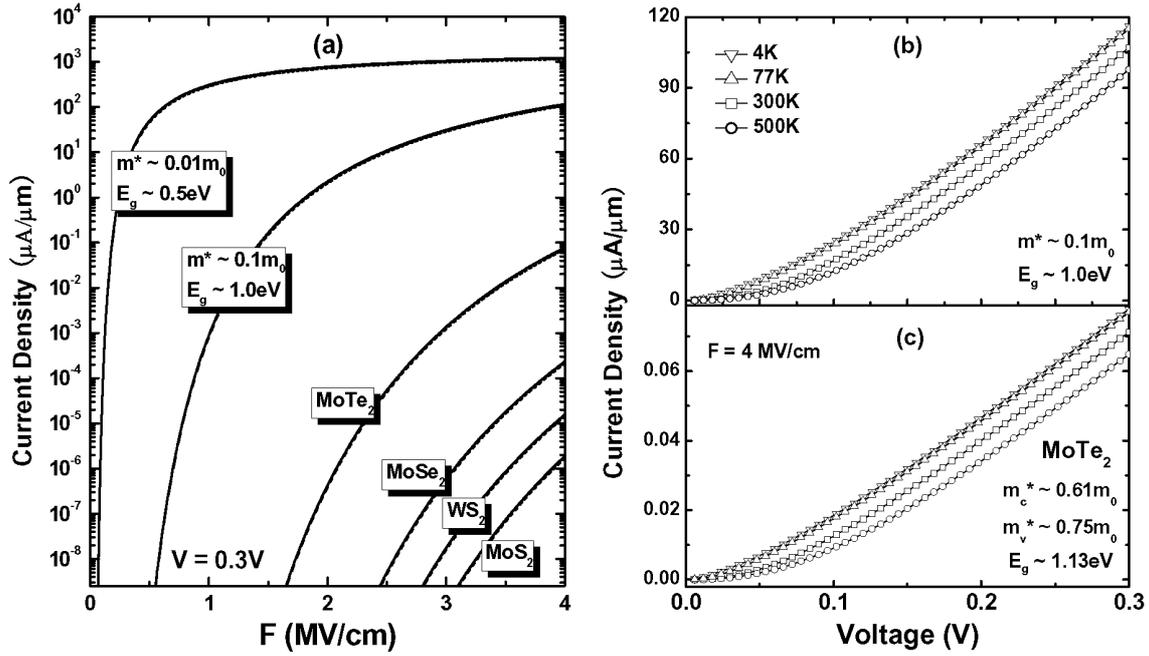

**Fig 2:** (a) Interband tunneling current density for various 2D crystal semiconductors. The solid lines are at $T$=4 K, and the dashed lines at $T$=300 K, the temperature dependence is weak. (b) Current-voltage curves at various temperatures at a junction field $F$=4 MV/cm for a 2D crystal semiconductor with band parameters indicated. (c) Same as (b), but for the 2D crystal MoTe$_2$.

The tunneling current densities for MoS$_2$ and the family of TMDs are found to be low owing to their large bandgaps. For example, the current density approaches ~0.1 μA/μm for MoTe$_2$ at a high field of 4 MV/cm. The tunneling current density of 2D graphene can be higher (~few mA/μm), but it lacks a bandgap. For TFET applications, 2D crystals with smaller bandgaps are necessary for boosting the current. For example, tunneling currents for 2D crystals with bandgaps of 0.5 eV and 1.0 eV with corresponding lower effective masses are plotted in Fig 2(a). The currents for such crystals exceed ~100 μA/μm at the highest junction fields, and thus can be attractive for high-performance TFET applications. Such small-bandgap materials could be intrinsic 2D crystals, or derived from interaction-induced bandgap of Dirac-cone surface states in thin topological insulator materials.[21] Another possibility is in bilayer graphene, where breaking the layer symmetry by vertical electric fields opens a small bandgap.

The tunneling current of MoTe$_2$ and a 2D semiconductor crystal with $E_g$ = 1.0 eV and $m^*$ = 0.1$m_0$ ($m_0$ is free electron mass) as a function of the voltage at different temperatures is shown in Fig. 2 (b) and (c). Note the $J_T^{2D} \sim V^{3/2}$ dependence at low voltages, the approximately linear relation $J_T^{2D} \sim V$ at high voltages, and the rather weak temperature dependence. Note that the current for the small effective mass 2D crystal is orders of magnitude higher than MoTe$_2$ in Figs 2 (b) and (c), even though their bandgaps are similar. Is this always true? Comparing the material parameters of Figure 2 (b) and (c), a natural question is the relative importance of effective masses and bandgaps. For III-V 3D semiconductors, the effective masses are proportional to the bandgaps, as would be expected from interband repulsive interaction from basic perturbation theory.[22] The equivalent picture is not clear yet for 2D semiconducting crystals. Therefore, we discuss all possibilities by treating the effective mass and bandgap as independent material parameters. Fig 3 shows the interband tunneling currents in 2D crystals at a high junction field for various bandgaps, plotted for a range of effective masses.

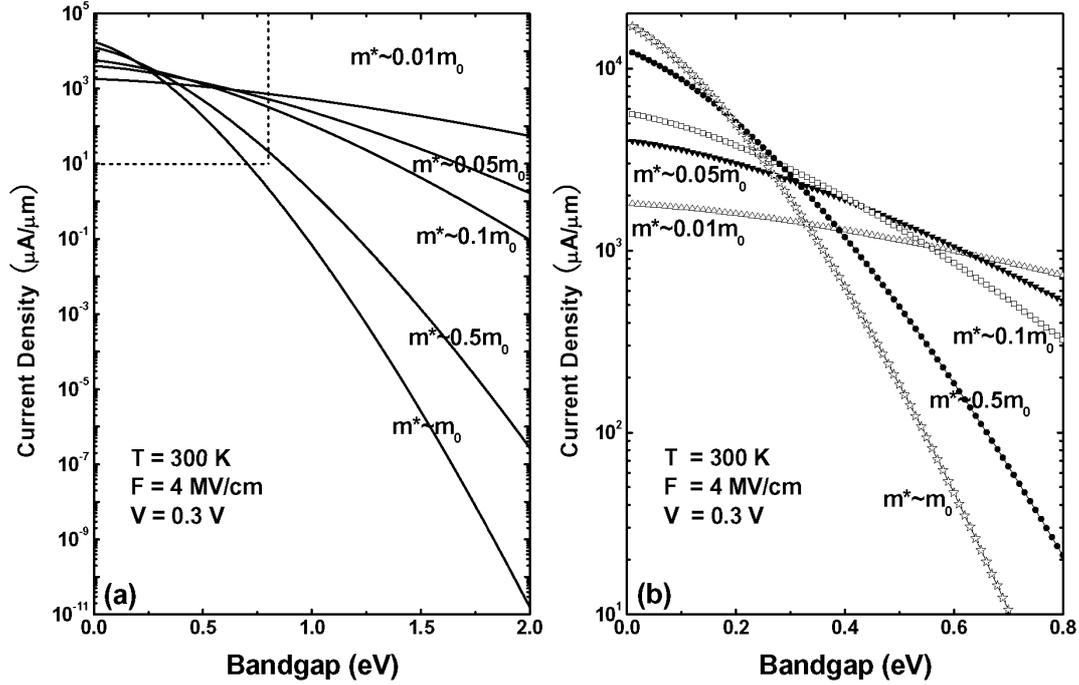

**Fig 3:** (a) Interband tunneling currents for 2D crystal semiconductors as a function of the energy bandgap for various effective mass parameters. (b)The high current part of (a) zoomed in for more details.

As is evident from Fig 3 (a), there is a tradeoff in the choice of effective mass and bandgap for maximizing the tunneling current. Fig 3(b) zooms in to highlight this crossover. For high-performance TFETs for digital switching applications, currents exceeding 100 μA/μm are highly desirable. For 2D crystals semiconductors with bandgaps smaller than ~0.3 – 0.4 eV, a choice of a *higher* effective mass will maximize the interband tunneling current, far exceeding typical transistor on-currents for high-performance switching. But for larger bandgaps, a *lower* effective mass is more desirable. It is essential that TFET devices *switch off*, which may be problematic for 2D crystals with very small bandgaps. For high-performance TFETs, 2D crystals with bandgaps in the ~0.6-0.7 eV range and effective masses of 0.1-0.5$m_0$ can thus be potentially very attractive.

On the other hand, there are various low-power applications of wider bandgap 2D crystals that do not require high on-currents, as long as they can beat the Boltzmann limit of 60 mV/decade. Such applications may include low-power sensors.[23] Other applications are expected to emerge as such devices become available. For such applications, the low tunneling currents of TMD 2D crystals might be actually turned into an advantage. We summarize this work by hoping that the analytical evaluation of tunneling currents in 2D crystals will be found useful for guiding the choice of the right materials for the specific applications. The analytic expression of the current vs. voltage can also form the backbone for compact modeling and design of 2D crystal TFETs when combined with the device electrostatics.

The authors acknowledge partial financial support from the NRI MIND center supported by the Semiconductor Research Corporation (SRC), the SRC/DARPA LEAST center, NSF, and AFOSR, and fruitful technical discussions with Prof. Alan Seabaugh.